\documentclass[a4paper, 11pt, twocolumn]{article}
\usepackage{natbib}
\usepackage{times}
\usepackage{graphicx}
\begin{document}

\begin{center}
{\Large\bf Explosive and radio-selected Transients: Transient Astronomy with SKA and its Precursors }\\
Poonam Chandra\footnote{National Centre for Radio Astrophysics, TIFR, Ganeshkhind, Pune 411007},
G. C. Anupama\footnote{Indian Institute of Astrophysics,II Block, Koramangala, Bangalore 560034},
K. G. Arun\footnote{Chennai Mathematical Institute, Siruseri, Tamilnadu 603103},
Shabnam Iyyani\footnote{Tata Institute of Fundamental Research, Dr. Homi Bhabha Road, Colaba Mumbai 400005},
Kuntal  Misra\footnote{Aryabhatta Research Institute of Observational Sciences, Manora Peak, Nainital 263002}, 
D. Narasimha$^4$,
 Alak Ray$^4$,
L. Resmi\footnote{Indian Institute 
of Space Science \& Technology, Thiruvananthapuram 695547}, 
Subhashis Roy$^1$,
Firoza Sutaria$^2$
\end{center}



\abstract

With the high sensitivity and wide-field coverage of the Square Kilometre Array (SKA), large
samples of explosive transients are expected to be discovered. 
Radio wavelengths, especially in commensal survey mode, are 
particularly well suited for uncovering the complex transient phenomena.
This is because observations
at radio wavelengths may suffer less obscuration than in other bands (e.g.
optical/IR or X-rays) due to dust absorption. 
At the same time, multiwaveband information often provides critical
source classification rapidly than possible with only radio band data.
Therefore,
multiwaveband observational efforts with wide fields of view will be the
key to progress of transients astronomy from the middle 2020s offering
unprecedented deep images and high spatial and spectral resolutions.
Radio observations of gamma ray bursts (GRBs) with SKA will uncover not only much fainter bursts
and verifying claims of sensitivity limited population versus intrinsically dim GRBs, they will also unravel  the
enigmatic population of orphan afterglows. The supernova rate problem caused by dust extinction in optical bands is
expected to be solved in the SKA era. In addition,  the debate of single degenerate scenario
versus double degenerate scenario will be put to rest for the progenitors of thermonuclear supernovae, since highly sensitive measurements
will lead to very accurate mass loss estimation in these supernovae. One also expects to detect gravitationally lensed supernovae in far away
Universe in the SKA bands.  Radio counterparts of the gravitational waves are likely to become a reality once SKA comes online. In addition, SKA is
likely to discover various new kinds of transients.


\section{Introduction}

Exploration of the transient Universe is an exciting and fast-emerging area within the radio astronomy.
Square kilometre Array (SKA) when ready will be world's largest radio telescope, 
to be built in Australia and South Africa. It will be built in two phases. 
Phase  I (SKA-1) is expected to   complete in 2020; 
Phase II (SKA-2) will be completed in 2024. Phase I will have  $\sim10$\% of 
the total collecting area.
SKA will have a range of detectors  including aperture arrays as well as  dishes, and  will span a frequency range from 
few tens of megahertz to a few gigahertz. SKA will have unique combination of great sensitivity, wide field of view and unprecedented computing power.
 With its wider field of view and higher sensitivity, the 
SKA, holds great potential to revolutionize this relatively new and exciting field, thereby opening up incredible 
new science avenues in astrophysics.

\subsection{Explosive Transients} 

  Transient radio sources are  compact sources and are 
locations of explosive or dynamic events.
Transient phenomenon usually 
represent extremes of gravity,  magnetic fields, velocity, temperature, pressure and density. 
In terms of duration, radio transients phenomenon can be classified into two classes, 
long time variability (min-days) and bursting radio sky \citep[msec-sec; ][]{carilli}. 
 Transients in radio bands are mainly dominated by three kinds of emission mechanisms:\\
 1:) {\bf Incoherent synchrotron emission:} 
Transients with incoherent emission show relatively slow variability,
and are limited by brightness temperature. These events are mainly
associated with explosive events, such as Gamma Ray Bursts (GRBs), 
supernovae (SNe), X-ray binaries (XRBs), tidal disruption
events (TDE) etc. 
Incoherent transients are mostly discovered in the
images of the sky.\\
2:) {\bf Non-thermal coherent emission}: Transients associated with coherent emission
show relatively fast variability, high brightness temperature and
often show high polarization associated with them, such as
Fast Radio bursts, pulsars, flare stars etc.
The coherent emission transients are
discovered in time series observations. The short-duration transients are 
   powerful probes of intervening media owing to dispersion, 
scattering and Faraday rotation. 
\\

3:) {\bf Thermal emission}: These kind of emission is 
seen in slow transients like novae and symbiotic stars.

\subsection{Radio selected transients}

A large fraction of
transients discovered in radio bands remain undetected in other
bands of electromagnetic spectrum and their nature remains unclassified. 
Early searches for variable and transients in radio band yielded only $\sim
10$ variable sources \citep{gt86,langston90}. Due to
lack of wide field monitoring of the transient sky, most of the variable radio
sources were found from follow up of X-ray transients and gamma-ray bursts.
However, with availability of wide field imaging techniques at low radio
frequencies and advancement in observing facilities allowed larger regions of
the sky to be monitored. In the last decade, use of these techniques have
resulted in several reported detections.

While at least 30 transients have been detected so far, the  radio detected transients typically have no identification in other
bands. This makes identifying the progenitors difficult. Out of the 30
transients detected, possible progenitor population of only about 9 have been
identified by the respective authors. These include two possible cases of
radio SNes (RSNes) or orphan GRBs \citep{bower07}, one confirmed SNe, two
confirmed low mass X-ray binaries, 3 probable scintillating AGNs and one
possible ultra-bright RSNe \citep{bannister11}. The origin of the rest of
the transients are not known. Clearly, this demonstrates that for most of the
transients detected first in radio band, the parent population is unknown.


\subsection{ Transients research in India and relevance to SKA}

 \begin{table*}
 \caption{Parameters for SKA-1 telescope}
   \label{tab1}
 \begin{tabular}{ccccccc}
 \hline
 Telescope &  Fiducial Freq. & FOV & Resolution & Baseline & Bandwidth & Sensitivity \\
                  & GHz & deg$^2$& arcsec & Km & MHz & $\mu$Jy-hr$^{-1/2}$\\
 \hline \hline
 SKA-1~LOW & 0.11 & 27 & 11 &  50 & 250 & 2.1\\
 SKA-1~MID  & 1.67 & 0.49 & 0.22 & 200 & 770 & 0.7 \\
 \hline
 \end{tabular}
 \end{table*}
 
In India active research is going on in transient astronomy, especially in the fields of supernovae, gamma ray bursts, Pulsars and related phenomenology, novae, X-ray binaries, tidal disruption events, Galactic black hole candidates, and electromagnetic (EM) signatures in gravitational waves. Major observational 
facilities operated by National Centre for Radio Astrophysics (NCRA), Aryabhatta Research Institute of Observational Sciences (ARIES), 
Indian Institute of Astrophysics (IIA), Inter University Centre of Astronomy \& Astrophysics (IUCAA) are widely used. 
Researchers also avail the wealth of archival data 
from high energy space missions like {\it{RXTE}}, 
{\it{XMM-Newton}}, {\it{Swift}}, {\it{Chandra}} and {\it{Fermi}}. Groups are also working towards multimessenger
 astronomy mainly in the context of gravitational waves. 
 Indian Astronomers have access to
several observational facilities, both presently existing and to
come in the future \citep[e.g., ][]{pandey}.

The dynamic radio sky is poorly sampled
as compared to sky in X-ray and $\gamma$-ray bands. 
To search for transient phase space, one needs  the combination of sensitivity, 
field of view and time on the sky. This has been difficult to realise simultaneously, hence
there have been a very few comprehensive transient surveys in radio bands. In addition,
the variation in time scales (nano seconds to minutes), and complex structures 
in frequency-time plane makes things further difficult. The advent of 
the multi-beam receivers placed on the single dish telescopes, such
as Parkes and Arecibo has led us to detect new phenomena such as rotating radio transients (RRATs), FRBs etc. However, much larger FOV is required to
discover more of such phenomena and SKA will be able to achieve desired sensitivity with large FoV (see Table \ref{tab1}).

SKA has adopted transients as its key science goal.
SKA will open up a new parameter space in the search for radio transients. 
Some of the basic questions to ask in context of SKA  are the following:
 What do we know about the radio
transient sky?
How can we improve searches for
radio transients?
What successes do we expect from
SKA pathfinders?
How will SKA pathfinders improve
our understanding of transients?

To achieve the above goals, one needs the right combination of 
different technologies. Many of these technologies and ideas are being tested 
 with the SKA pathfinders and precursors, such as Australian Square Kilometre Array Pathfinder (ASKAP), 
 Murchison Widefield Array (MWA), Low-Frequency Array (LOFAR), 
 upgraded-Giant Metrewave Radio Telescope (uGMRT) etc.
The use of widely distributed many small elements  for the SKA means that 
one  needs to take advantage of vast computing resources to be able to 
sample a larger fraction of the sky.
Also the discovery of truly transient events relies on having 
excellent instantaneous sensitivity. 
Here it is important to note that multiwavelength observations of the sky are 
important for the detection and follow-up of transient sources, as they 
provide complementary views of the same phenomena. 
The SKA will be roughly contemporaneous with other International facilities 
like Large Synoptic Survey Telescope (LSST), Gaia, Thirty Meter Telescope (TMT) 
in optical/IR bands and next generation successors of
{\it Swift}, {\it Chandra} and {\it XMM-Newton} in the gamma-ray and X-ray bands (such as {\it SVOM} (France-China), 
{\it SMART-X} (US) and {\it ATHENA} (ESA) missions). 

For SKA to be an optimal telescope for transient science, the
international transient science working group (SWG) has recommended mainly two changes to SKA
Phase 1 design:

{\bf Commensal Transient Searches:}\\
It is important to do near-real-time searching of
all data for transient event. 
This will increase  rate of events by at least one order of magnitude.
For maximum scientific gain,  the detection needs to be reported widely, globally and rapidly.
There are efforts going on to implement real-time transient pipeline in uGMRT to enable commensal transient searches. 

{\bf Rapid (robotic) Response to Triggers:}\\
One may be able to detect very
early-time radio emission (coherent or very prompt
synchrotron) if SKA has Robotic response to external 
alerts, allowing very rapid
follow-up of high-priority events.
For example, the Arcminute Micro-kelvin Imager Large Array (AMI-LA)
is world's first robotic automated radio telescope.
Its response time to {\it Swift} triggers are only 30 seconds. 
AMI-LA has provided the first millijansky-level constraints on prolonged radio emission from GRBs within the first hour post-burst \citep{staley}.

SKA is expected to increase number of transients 
at least by an order of magnitude.
In optical bands, LSST, which will be contemporary of SKA, anticipates finding around 1 million
transients per night. However, this rate is not known at the GHz bands. 
But it is estimated that 
around 1\% of the mJy sources are variable \citep{frail2012}.
LOFAR has recently detected a transient at 10 Jy flux density level 
at 60 MHz band \citep{stewart}. The transient does not repeat and has no counterpart. The nature of this transient is enigmatic. SKA~LOW will most likely detect 100s of such transients per day in MHz band.
\citet{frail2012} claims that with 10 $\mu$Jy rms, one can expect around
one transient per degree. This is achievable by SKA-1~MID.

\section{Gamma Ray Bursts}

Gamma Ray Bursts (GRBs) are non-recurring bright flashes of $\gamma$-rays lasting from seconds to minutes with an average observed rate of around one burst per day. The energy isotropically released in $\gamma$-rays ranges from $\sim 10^{48}$ to $\sim 10^{54}$~ergs. The prompt emission in majority of cases is in $\gamma$-ray bands. In a few cases, optical prompt emission is detected. The prompt emission spectrum is mostly non-thermal, though presence of thermal or quasi-thermal components are seen in a handful of bursts (see \cite{Kumar:2014upa} for a review). Afterglow emission following the burst in longer wavelengths and lasting for longer time was eventually discovered leading to the confirmation that the bursts are cosmological.   GRBs are one of the furthest known objects so far, the redshift distribution ranges from $0.008$ to $\sim 9$ as of now. We currently understand them as catastrophic events generating a central engine which launches an ultra-relativistic collimated outflow.

After nearly 4 decades of extensive research, even though our knowledge about GRBs have definitely enhanced, the 
very nature of the progenitor, and the radiation physics giving rise to the observed emission in both prompt as well afterglow phase is yet to be solved. This is mainly attributed to its nonrecurring behaviour, with the emission over only a short timescale resulting 
in unique and variable light curves. Some of the prominent questions that need to be answered are: what is the nature of the content of the outflow, is it baryonic dominated or Poynting flux?; what is the radiation physics: photospheric emission 
or non-thermal processes such as synchrotron emission or inverse Compton?; from where exactly in the outflow does the emission occur?; what are the microphysics involved in these radiation events?; what are the dynamics of the outflow resulting in these variable light curves?; how does the transition between the prompt and afterglow phase occur and what is its nature giving rise to the X-ray flares, 
plateaus, steep decay in flux observed in the early X-ray afterglow?; GRB are believed to be collimated jet emissions to reconcile with the observed energetics, however, this raises question as to why jet breaks are not observed in all afterglow observations?

Resolving the above queries would enable us to use GRB as an effective cosmological tool to probe the early universe. There are several theoretical models proposed for both prompt as well afterglow emission of GRBs which are sometimes equally consistent with the data. Therefore, we are at such a juncture where it is imperative for the forthcoming observations and studies of GRBs to be able to break the current degeneracy between existing models. 
Afterglow emission has always been instrumental in diagnosing burst physics like the energy in the explosion, structure of jet, nature of the ambient medium and the physics of relativistic shocks.
In this direction, it is evidently clear that we require a multi-wavelength observational coverage of these events extending from gamma rays to radio
bands, which can prove highly constraining for existing theoretical models. This in turn would then become effective in developing a global model connecting prompt to afterglow emission, which can also be adequately tested.  SKA, with its broad field of view and higher sensitivity would be able to play a key role in achieving this prospect as well as in resolving many of the above mentioned open questions.

\subsection{Long versus Short GRBs}

The histogram of GRB duration reveals a bimodal distribution \citep{Kouveliotou:1993yx}: the long and the short bursts. In the BATSE burst population, bursts with duration less than $2$~s were classified as short and those lasting for more than $2$~s were classified as long. Apart from the duration, long and short bursts also show difference in the hardness of their $\gamma$-ray spectrum \citep{Kouveliotou:1993yx}. Subsequently, long GRBs at lower redshifts are found to be associated with type Ib/c broad lined supernovae, thereby establishing their origin in the gravitational collapse of massive stars \citep{Woosley:2006fn}. No short burst has so far been found associated with supernovae. Further studies have revealed that the difference lies deeper than the prompt emission, and extends to the burst environment. In accordance with the massive star hypothesis, long GRBs are predominantly found in star forming regions of late type galaxies \citep{Fruchter:2006py} while short bursts are seen in a variety of environments \citep{Fong:2009bd}. Distribution of the off-set of burst position from the photocentre of the host galaxy shows that short bursts typically are further away from the central regions of the galaxy compared to their longer cousins \citep{Fong:2009bd}. These evidences support a hypothesis that long bursts are possible end stages of massive stars ($M > 15 M_{\odot}$ ) while short bursts are associated with older stellar populations like compact objects. Currently we believe that at least a fraction of short GRBs form due to the merger of compact object binaries (see \cite{Berger:2013jza} for a review). Recently a new class of GRBs, named ultra-long bursts (ULGRBs), are found to be lasting for several thousands of seconds \citep{Levan:2015ama}. It is speculated that ULGRBs are arising from a different class of progenitors.
With the recent discovery of gravitational waves (GWs), a new era of Gravitational Wave Astronomy has begun which can shed more light into this model if short GRBs or their afterglows are observed in coincidence with double NS or NS-BH mergers. (See Sec 5  of \citet{ska} for capabilities of SKA to carry out electromagnetic follow up of GW events).

The differences in various progenitors giving rise to different kinds of GRBs will be manifested in their environments. However, in X-ray and optical bands GRB, fireball is almost always optically thin barring a few cases. The early evolution of low frequency ($< 4$~GHz) radio lightcurves is through the optically thick regime, the lightcurve peak corresponds to transition from optically thick to thin regime. Hence, radio frequencies are unique in probing the evolution of the self-absorption frequency, $\nu_a$, which in turn can constrain the physical parameters such as density of the environments etc.. Tracing the evolution of 
$\nu_a$ will also help in scrutinizing basic assumptions in the standard fireball model, especially of the time-invariant shock microphysics.
SKA can  densely sample the lightcurves of long as well as short GRBs at
different frequencies. 
Thus radio emission from long versus short GRBs will reveal the
difference in the environments of these two classes of GRBs.

 In the {\it Swift} era, almost 93\% of GRBs have a detected X-ray afterglow, $\sim$ 75\% have detected optical afterglows while only 31\% have radio afterglows \citep{Chandra:2011fp}. The SKA, with its extreme sensitivity, will dramatically improve statistics of radio afterglows. Using numerical simulation of the forward shock, \cite{Burlon:2015yqa} predicts around $400 - 500$  sr$^{-1}$ yr$^{-1}$  radio
 afterglows in SKA-1~MID bands. 
However, \citet{hancock} has claimed  that undetected afterglows are intrinsically dim rather than sensitivity limited.
SKA will be very crucial in lifting this 
degeneracy owing to its extremely high sensitivity.
 \citep{Burlon:2015yqa} have shown that the  
band 5 of SKA1-MID is ideally suited to observe  radio afterglow 
during first 10 days, when the afterglow emission will be 
closest to the peak (Table \ref{tab1}).

Both long and short GRBs are collimated events with opening angles lying within few degrees. 
This means we can detect only those GRBs which are pointing towards us. 
However, in the absence of $\gamma$-ray trigger, either due to the jet pointing away from earth or due to unavailability of a dedicated triggering instrument, orphan radio afterglows can be detected by SKA owing to its high sensitivity. \cite{Burlon:2015yqa} predicts $300 - 400$ orphan afterglows per all sky per week. Orphan afterglow detection rates will help in drawing indirect limits of jet structure, as well as true rate of GRBs.

Thus SKA will trace the peak of the emission, sampling the
 the evolving spectrum, and follow-up the afterglow  to non-relativistic regime.  SKA-1~MID Bands 4 (4 GHz) and 5 (9.2 GHz) will be particularly suited for GRB afterglow observations.

\subsection{Unveiling the GRB Reverse Shock with SKA} 

According to the standard model, a central engine, perhaps a black-hole torus system or a millisecond magnetar, formed during either the gravitational collapse of a massive star (for long GRBs) or the merger of a double compact object binary (short bursts), launches an ultra-relativistic jet. The jet energy could either be in the bulk kinetic energy of its baryonic content or in magnetic fields (Poynting Flux Dominated Outflow). Prompt emission arises due to internal dissipation of the  outflow energy. The remaining energy gets dissipated in the medium surrounding the burst ensuing longer wavelength afterglow emission. However, in majority of GRBs, the early afterglow is influenced by continuous energy supply from the central engine.

The external energy dissipation happens through a forward and reverse shock system. The forward shock runs into the ambient medium and the reverse shock runs into the ejected material. Emission from the reverse shock is seen in the early optical \citep{Sari:1999iz} and radio afterglows \citep{Laskar:2013uza}. Early afterglow is one of the main tools for probing the central engine. Particularly, the reverse shock emission can shed light into the magnetization of the ejecta which is an important parameter in knowing the jet launching mechanism.

\begin{figure}
    \centering
    \includegraphics[width=0.45\textwidth,natwidth=610,natheight=642]{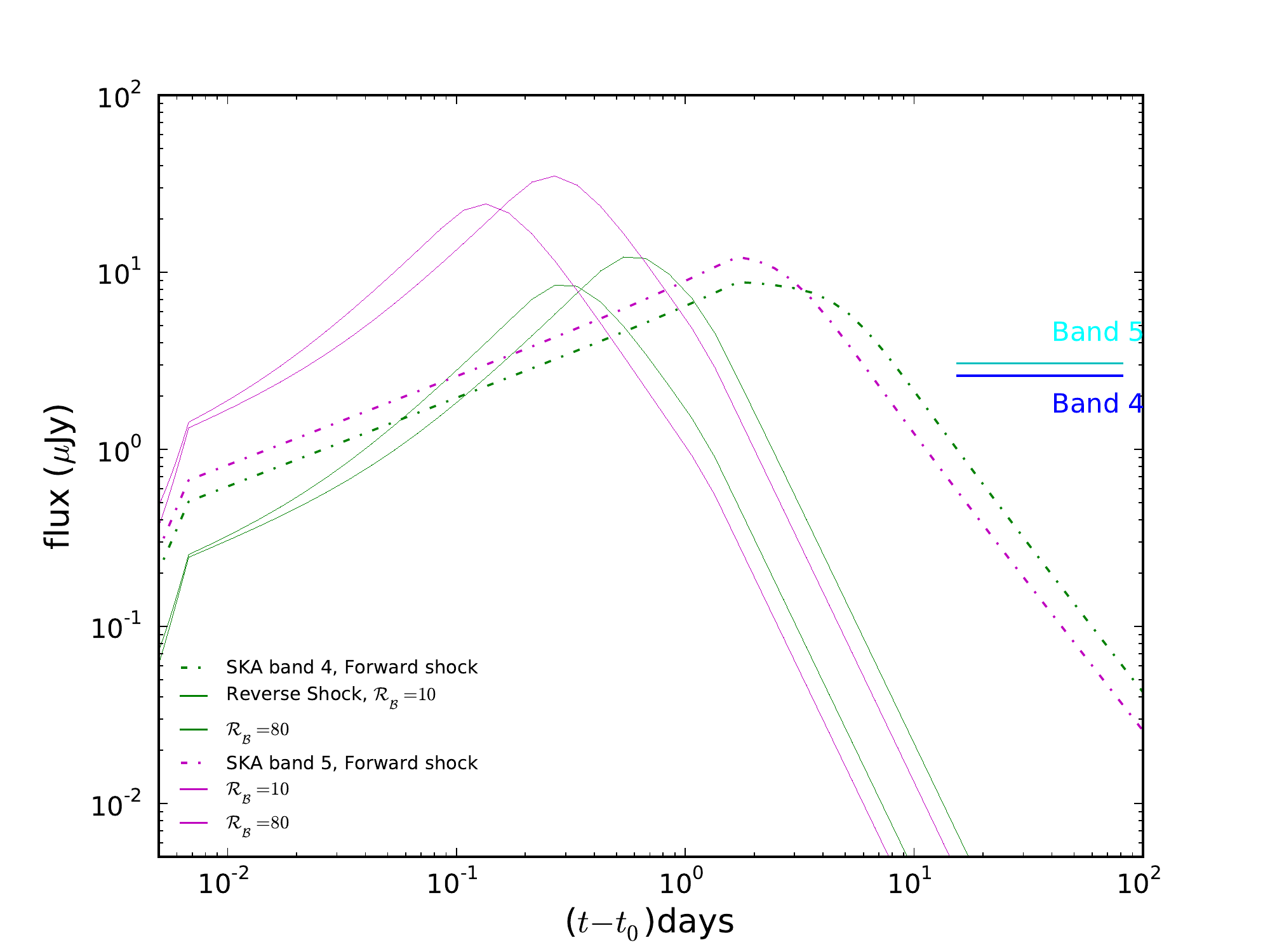}
    \includegraphics[width=0.45\textwidth,natwidth=610,natheight=642]{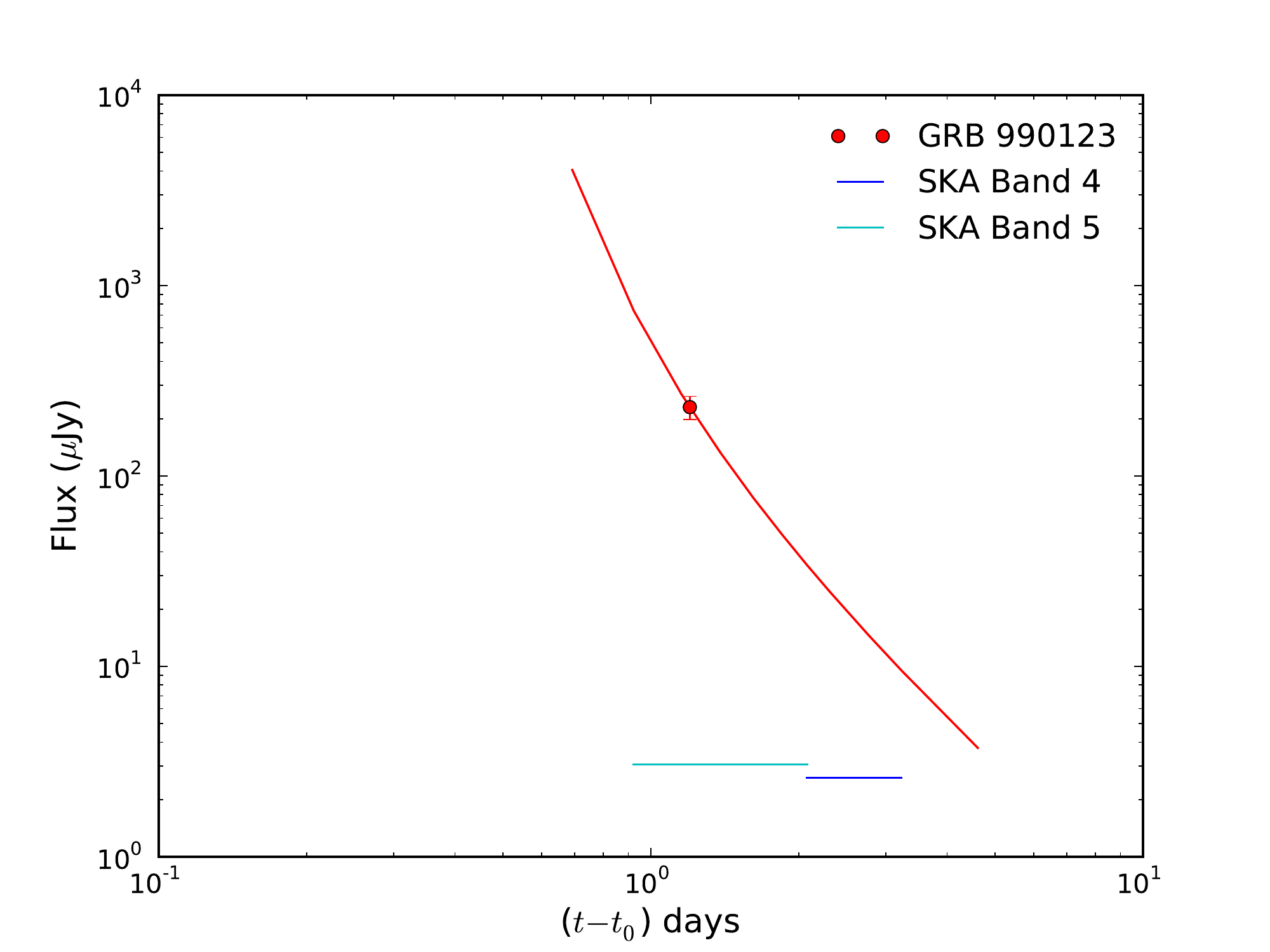}
\caption{(Top): Reverse and forward shock predictions in SKA-1~MID Band-4 ($4$ GHz) and Band-5 ($9.2$ GHz) for standard parameters (energy in explosion $E_{\rm iso} = 10^{52}$ergs; ambient medium density $n_0 = 0.1$ atom/cc; initial bulk Lorentz factor $\eta = 100$; energy content in forward and reverse shock electron populations, $\epsilon_e = 0.1$; energy content in forward shock magnetic field $\epsilon_B = 10^{-5}$). The burst is assumed to be at a redshift of $2$. Reverse shock lightcurve strongly depends on the ratio $\cal{R}_B$ of reverse to forward shock magnetic field energy density, in turn the magnetization of the ejecta. $3 \sigma$ flux limits for SKA~MID Band-4 and Band-5 for a $2$hr on source integration are also shown for comparison. (Bottom): Radio flare of GRB$990123$ in VLA $8$ GHz scaled to different redshifts. $3 \sigma$ limits of SKA-1~MID Band-4 and Band-5 are plotted in the relevant redshift ranges.}
\label{fig1}
\end{figure}

There have been several observational as well as theoretical studies of radio reverse shock emission in the literature. The first detection of a radio flare, expected to be originating from the reverse shock, was for GRB990123. In recent times, deep and fast monitoring campaigns of radio reverse shock emission could be achieved using JVLA \citep{Chandra:2011fp, Laskar:2013uza} for a number of bursts. Reverse shock emission is brighter in higher radio frequencies where self-absorption effects are relatively lesser. \citet{Gao:2013mia, Kopac:2015dia} and \citet{Resmi:2016a} have done comprehensive analytical and numerical calculations of radio reverse shock emission. Figure \ref{fig1}  compares theoretical predictions of reverse shock emission from \citet{Resmi:2016a}. The lightcurves are calculated by assuming a Newtonian reverse shock (thin shell regime) and ultra-relativistic forward shock system in a constant density ambient medium. Difference in ejecta magnetization is reflected in the RS lightcurve peak time. Radio afterglow monitoring campaigns in higher SKA bands will definitely be useful in exploring reverse shock characteristics. In Figure \ref{fig1}  we look into the detectability of GRB990123 had it been at higher redshifts. The $8$ GHz peak flux and corresponding $t_{\rm obs}$ are scaled accordingly. Accounting for cosmological k-correction, the flux will correspond to SKA-1~MID Band-4 and Band-5 as the redshift changes. SKA-1~MID will be able to detect  a bright radio flare like GRB990123 even if it happens at a redshift of $\sim 10$.

\subsection{Late afterglow and calorimetry with SKA}

\begin{figure}
    \centering
    \includegraphics[width=0.45\textwidth,natwidth=610,natheight=642]{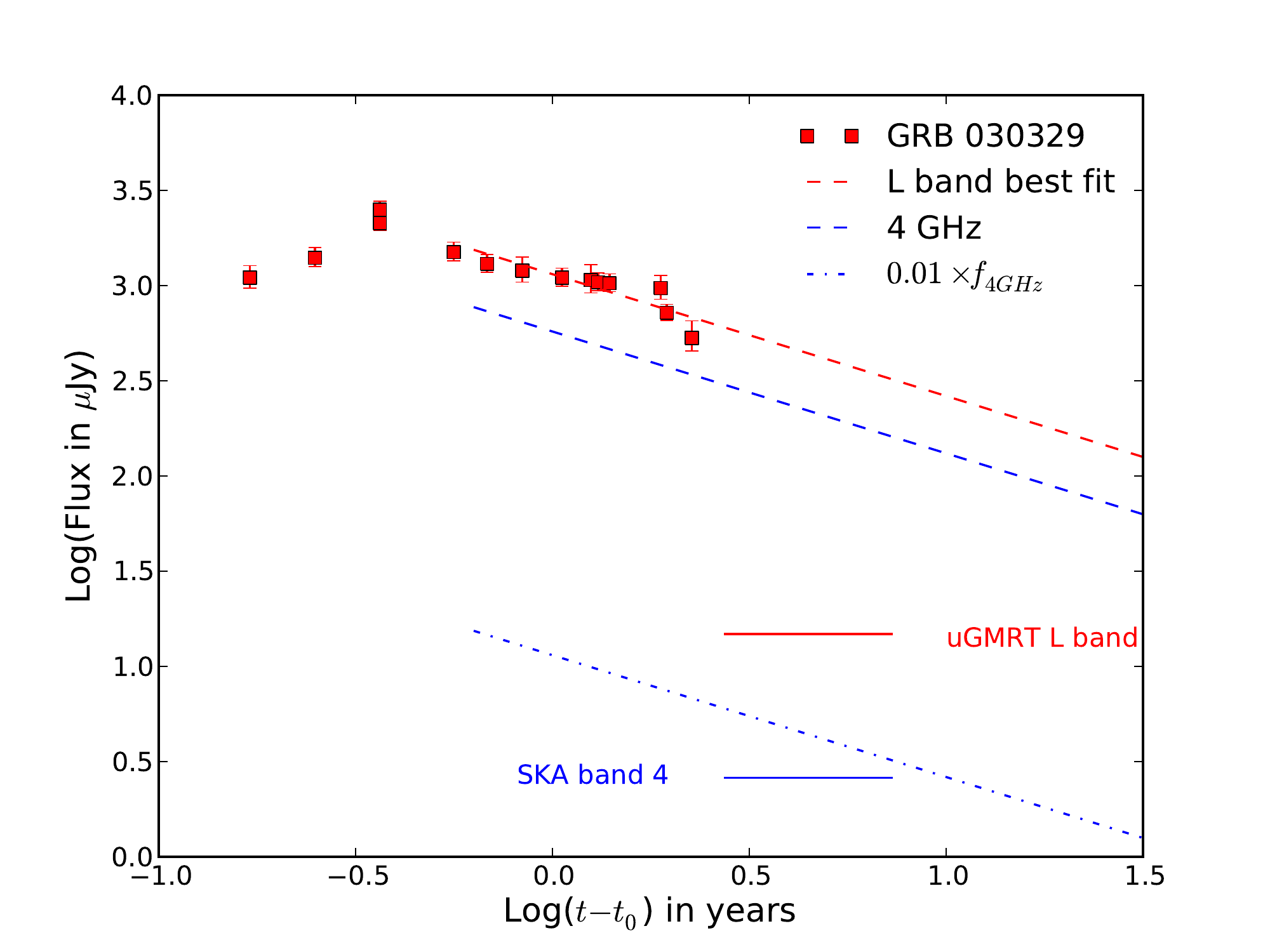}
\caption{GMRT L-band lightcurve of GRB030329. Best power-law fit to late time L-band data is shown as dashed red line. A spectral factor assuming index $-0.5$ is used to obtain the $4$~GHz extrapolation (blue dashed line). uGMRT and SKA Band-4 flux limits are for $2$hr on source time. The dash-dot line correspond to a burst $100$ times fainter than GRB030329 in $4$GHz.}
\label{fig2}
\end{figure}

The uncertainty in jet structure in GRBs pose difficulty in constraining the energy budget of GRBs.
Following the afterglow till non-relativistic transition is the key to constrain the GRB energetics independent of  
geometry.
While X-ray and optical afterglows stay above detection limits only for weeks or months, radio afterglows of nearby bursts can be detected upto years \citep{Frail:1999hk,Resmi:2005bj}.  A couple of very bright afterglows have been observed in time scale of several years by VLA and GMRT.  GRB 030329 is the longest observed afterglow, in a campaign made possible through GMRT low frequency capabilities even when the afterglow had gone below detection limits in higher frequencies. In figure-\ref{fig2} we plot the first two years of observations of GRB030329 radio afterglow in GMRT L band.

The longevity of radio afterglows also make them unique laboratory to study the dynamics and evolution of relativistic shocks. Fireball transition into non-relativistic regime can be probed through low frequency radio bands \citep{Frail:1999hk,vanderHorst:2007ir} However, this regime is largely unexplored due to limited number  of bursts in past that stayed above detection limit beyond sub-relativistic regime. Several numerical calculations exist for the afterglow evolution starting from the relativistic phase and ending in the deep non-relativistic phase \citep{vanEerten:2011bf,DeColle:2011ca}. SKA with its $\mu$Jy  level sensitivity will be able to extend the afterglow follow up time scale. This will provide us with an unprecedented opportunity to study the deep non-relativistic regime of afterglow dynamics and thereby will be able to refine our understanding of relativistic to non-relativistic transition of the blastwave as well as  shock microphysics in the GRBs. 
In the SKA era, these predictions can be tested against observations. Late afterglow phase where the fireball is both sub-relativistic and nearly spherical, calorimetry can be employed to obtain the burst energetics. These estimates will be free of relativistic effects and collimation corrections.
  \citet{Burlon:2015yqa} have computed the rate of detectable 
GRB afterglows in non-relativistic regime to be 
about 25\% with full SKA.

\subsection{Connecting prompt and afterglow Physics}

GRB prompt spectra in general look non-thermal in nature, and is generally modelled using a Band function, 
which is a phenomenological function of two power laws smoothly joined at a peak parameterised as $\rm E_{\rm peak}$ 
\citep{Band1993}. However, the study of GRB spectra using non-thermal models pointed out several spectral features as well as questions 
\citep{Preece1998,Goldstein2013,Gruber2014, Amati2002,Lu2010,Amati2009} and more importantly, the origin of the observed high radiation efficiency in gamma rays \citep{Cenko2010,Racusin2011}.
Pure non-thermal models faced difficulty in addressing these questions, which evoked the idea of the presence of thermal component, which is inherent in the standard fireball model, along with non-thermal component in the spectrum.  Theoretical models of photospheric emission have currently become the pivot of the study of prompt emission in GRBs. 
The time resolved analysis of the spectrum enables us to follow the dynamic spectral evolution as well that in these outflow parameters with time, which is crucial to our understanding of how progenitor evolves within the burst \citep{Iyyani2013,Iyyani2015,Iyyani2016}.

BATSE  was limited due to its insufficient energy window (20 $-$2000 keV) in determining the overall shape of the GRB spectrum. The {\it Swift} hard X-ray detector BAT, however, is not very sensitive above about 150 keV  \citep{Meegan09, Band2006}  and hence cannot measure the prompt X-ray/ gamma-ray spectrum accurately (majority of GRBs have their peak energy around 250 keV). 
However, this drawback was overcome by the launch of {\it Fermi} in 2008, providing observation over a broad energy range of nearly 7 orders of energy, 8 keV - 300 GeV, made possible by two main instruments onboard: Gamma ray burst monitor  \citep[GBM; ][]{Meegan09} and Large Area Telescope  \citep[LAT; ][]{Atwood09}.
Some of the key observations by {\it Fermi} had been: i) the delayed onset of high energy emission for both long and short GRBs \citep{Abdo2009_nature,Abdo2009_080916c,Abdo2009_090902B}, ii) Longer lasting LAT emission  \citep{Ackermann2013}, iii) very high Lorentz factors ($\sim 1000$) are inferred for the detection of LAT high energy photons \citep{Abdo2009_nature}, iv) several bright bursts show a significant detection of multiple emission components such as thermal component \citep{Guiriec2011,Axelsson2012,Burgess2014a}, power law \citep{Abdo2009_090902B} or spectral cut off at high energies \citep{Ackermann2011}, in addition to the traditional Band function, iv) interestingly, the Band function is found to have a different spectral shape in these multi-component fitting \citep{Guiriec2013} and v) recently, it was shown by \citet{Axelsson2015} that $78 \%$ of long GRBs and $85 \%$ of short GRBs have spectra with spectral width at full width half maximum of the $\nu \rm F_{\rm \nu}$ peak, that is much narrower than optically thin synchrotron emission, but still broader than a blackbody.

    While {\it Swift} and {\it Fermi} satellites have enriched our understanding of the GRBs by wide band spectral measurement during the prompt emission by the {\it Fermi} satellite and quick localisation, follow up observation and the consequent redshift measurement by the {\it Swift} satellite \citep{geh09, geh12}.
 But currently good spectral and timing measurement from early prompt phase
to late afterglow is available for a few sources and they are particularly lacking for the short GRBs.
Some of the key problems that can be addressed by the observation of the radio afterglows by SKA in connection with the prompt emission would be the following:\\
i) The connection of LAT detected GRBs with delayed high energy emission with afterglow parameters and comparing the Lorentz factor estimation with both LAT detected GeV photons as well as from the afterglow physics \citep{Kidd2014}. If the photosphere emission is detected in the prompt phase, this will also give an estimate of the dynamics of the outflow \citep{Iyyani2013}. This effort would enable to constrain the Lorentz factor of the outflow which is one of the key dynamic parameters directly connected to the progenitor.\\
ii) Evaluating the connection as well as comparison between non-thermal emission of both the prompt as well as afterglow emission. This would enable to constrain the microphysics of the shocks accelerating electrons to ultra-relativistic energies eventually producing the observed radiation.\\
iii) Detailed modelling of the afterglow observation of both long and short GRBs, will enhance our knowledge about the circumstellar as well as interstellar medium surrounding the progenitors. This will provide information about the host galaxy and the environment which will be crucial in assessing the various proposed progenitor models for these bursts.

SKA with its finer sensitivity would play a key role in constraining the energetics of GRBs which is crucial in estimating the radiation efficiency of the prompt emission of GRBs. This would include the crucial detection of jet breaks in radio bands as well as in constraining the dynamics of the outflow parameters. This would also strengthen our understanding of the hardness - intensity correlation \citep{Amati2002} which can eventually enable us to adopt GRBs as standard candles to probe cosmology. Till to date GRB is the most distant event probed by any astronomical observatory at a photometric redshift estimate of z = 9.4 detected for the burst GRB 090429B \citep{Cucchiara2011}.
In co-ordination with both space as well as ground based observatories in other wavelengths, SKA observations can be crucial in studying the timing of the onset of emission in various wavelengths which can throw light on the issue of where in the outflow does the emission occurs.
There have been several detections of thermal component in early X-ray afterglow \citep{Campana2006,Starling2012,Friis&Watson2013}. It would be interesting to find the component in longer wavelengths extending till radio bands which can then prove highly constraining in assessing its connection to the observed prompt thermal emission or whether it has a completely different origin.

The recently launched    {\it Astrosat}  satellite \citep{singh14} carries a suite of instruments  for 
multi-wavelength studies and   the Cadmium Zinc  Telluride  Imager (CZTI) 
can act as a monitor above about 80 keV \citep{rao15, bhalerao15}.  CZTI, along with {\it Swift}, can provide prompt spectroscopy and
CZTI can localise {\it Fermi} detected GRBs correct to a few degrees. Further, for bright GRBs it can
provide time resolved polarisation measurements. Hence, for a few selected bright GRBs, CZTI, in 
conjunction with other operational satellites, can provide wide band spectra-polarimetric information
and if followed up by other ground based observatories like SKA pathfinder uGMRT, we can have a complete observational
picture of a few bright GRBs from early prompt phase to late afterglow. This will provide us with
a comprehensive picture of GRBs, thus enabling a good understanding of the emission mechanisms.

\section{Supernovae}

Supernovae (SNe) are explosive events with two basic types: 1. thermonuclear Supernovae (SNe-Ia), caused by the explosion of a massive  white dwarf in a binary system. Their
optical spectra are characterized by the absence of H lines and the presence of  Si II absorption line; 2. Core-Collapse Supernovae (CCSNe), 
which mark the end
of the life of massive stars.  Their classification is based on the observed spectra. Presence or absence of Hydrogen lines classify
them as Type II versus Type I. In Type I SNe, presence or absence of silicon lines categorize them between Type Ia versus Type Ib/c. Further
Type Ib and Ic are classified based on the presence and absence of Helium lines, respectively. Amongst Type II SNe, the varied nature of
light curves classify them into Types IIP, IIL and IIn.

The physical processes driving the radio emission, and its temporal and spectral properties, turns out 
to be not only a characteristic of these explosive events itself, but of the pre-explosive evolutionary history of their precursor (progenitor) as well. Typically, there is almost no thermal radio emission in these events. A common property of all supernovae which show up in the radio is the presence of some amount of circumstellar medium (CSM), either or diffuse, in the vicinity of the progenitor, and the synchrotron emission is often successfully modelled as being produced by the interaction of the explosion shock with this material. Incoherent radio emission  has been detected from various types of core-collapse supernovae, which includes SNe-Ib  \citep[e.g. 
PS15bgt ][]{kamble15}, SNe-Ic \citep[e.g. SN 2012ap, ][]{chakraborti15},   SNe-IIn, e.g. SN 1995N \citep{chandra09}, SN 2006jd \citep{chandra12}, SN 2010jl \citep{chandra15}, and 
SNe-IIP  \citep[e.g. SN 2004et, ][]{misra07},
 2011ja \citep{chakraborti13}, 
20012aw \citep{yadav14,chakraborti16}.

Ccommensal SKA observations have potential to discover all CCSNe in the local universe, thus yielding an accurate
determination of the volumetric CCSN rate.
SKA will also be contemporaneous with 
LSST and WISE,  thus opening a radio vista in to the statistical properties of a dynamic universe, while as a follow-up
instrument, its high sensitivity and resolution will answer several crucial questions about the nature and origins of 
various SNe, both in the early and in the present Universe. 

\subsection{Radio view of Core Collapse Supernovae}

In supernovae, radio emission arises from shock interaction with the surrounding  circumstellar medium (CSM), which  provides
an important probe to see their last evolution stage.
Since radio-SNe can show both early and late radio 
emission, the most probable model would be that the emission itself comes from relativistic electrons trapped in shocked CSM (early emission) or from the 
reverse shock itself, as it propagates inwards in to the SNe ejecta (late emission)\citep{chevalier98} Therefore, radio SNe  probe the 
dynamics of the SN ejecta, the progenitor structure, and the CSM. 
Early radio emission is absorbed by either due to free-free absorption (FFA) or synchrotron self absorption (SSA).  Since SSA necessarily requires the presence of a magnetic field in the emitting region, 
radio spectra also provide insight in to the nature of the CSM field, as well as in to the generation/disruption of fields in the SNe-ejecta. 
In addition, signatures
in the multifrequency radio light curves give concurrent information about how the shock accelerated 
electrons cool in the presence of a long lasting optical emission from the SN through Inverse Compton (IC) cooling or
instead, whether synchrotron cooling dominates
over IC cooling \citep{chevalier06}.
 Compton cooling of electrons could keep
the radio light curve relatively flat in the optically thin regime, whereas synchrotron cooling of electrons will be important if the 
 magnetic field is efficiently built up in the explosion.
Radio light curves around 100 days can indicate whether Compton cooling dominates over
synchrotron cooling \citep[Fig. 5 of ][]{chevalier06}.

\begin{figure}
\centering
\includegraphics[width=0.35\textwidth,natwidth=610,natheight=642]{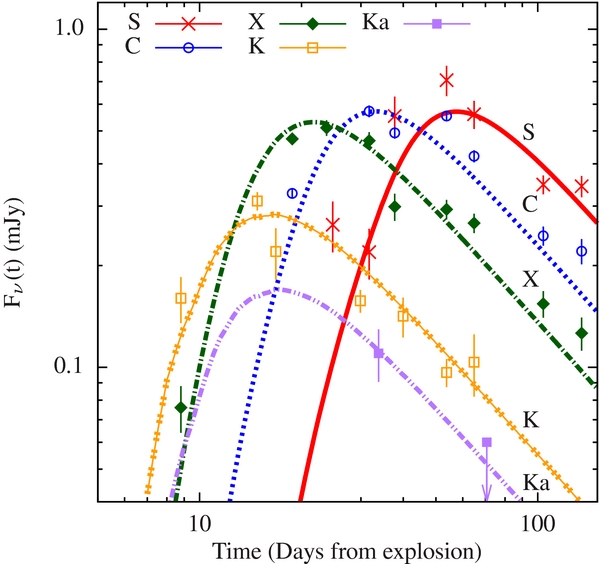}
\caption{Radio observations of SN 2012aw. The bands are as follows: S (3.0 GHz), C (5.0 GHz), X (8.5 GHz), K (21.0 GHz), and Ka (32.0 GHz) \citep[reproduced from ][]{yadav14}.}
\label{fig3}
\end{figure}

Radio data of a few type IIP SNe show long periods of radio brightness (e.g. beyond 100 days as in SN 2004et, Fig/ \ref{fig3}).
Radio data on
SN 2012aw shows the rapid evolution of the radio spectrum between 30 to 70 days and a signature of cooling at higher frequencies \citep{chevalier06}. Early observations at high frequencies are critical to the issue of IC cooling of radio
emitting electrons. A fraction of type IIP supernovae e.g. 2011ja \citep{chakraborti13}
may happen inside circumstellar bubbles blown
by hot progenitor or with complex circumstellar
environments set up by variable winds. 

SNe IIn are most complex in terms of stellar evolution.
They encompass
objects of different stellar evolution and mass loss history.
Some SNe IIn, like 
SN 2005gl, appear to have a luminous 
blue variable (LBV) star as their progenitor 
\citep{galyam07}.
 A special class of SNe IIn is characterized
by spectral similarities to SNe Ia at
peak light, but later shows SN IIn properties, e.g., SNe 2002ic and 2005gj.
Some SNe IIn appear to be related to SNe Ib/c. 
Most recently SN 2009ip is the unique Type IIn SN to have both a massive blue
progenitor and LBV like episodic ejections \citep{mauerhan12,margutti13}.
 A challenging problem in SNe IIn is that 
 their evolutionary status, as well as the
origin of the tremendous mass loss rates of 
their pre-SN progenitor stars is known. The estimated mass-loss rates of SN IIn progenitors are much higher than those of usual
massive stars ($10^{-4}-10^{-1}$ $M_\odot \rm ~yr^{-1}$). Thus SN IIn progenitors must have enhanced mass loss
shortly before their explosions. 
Thus to unveil the nature of the mysterious SN IIn
progenitors and mass loss related to them, we need to know the properties of the SN ejecta
and dense CSM related to them.

Some SNe-Ib/Ic have been associated with long duration GRBs. However, as the cases of SN2009bb and SN 2012ap have shown, there may be a class of 
stellar explosions which bridge the continuum between jetted relativistic events like GRBs and the more spherically symmetric non-relativistic "garden 
variety" supernovae, without high energy counterparts. These recently discovered radio bright hypernovae (R-HNe) raise important questions about the 
nature of their central driving engines. The possibilities include: (a) relativistic jets which are formed  deep inside the progenitor stars (and whose 
emergence is suppressed) \citep{margutti14}; (b) quasi spherical magnetized winds from fast-rotating magnetars, leading to 
asymmetric explosions, and (c) re-energising of the shock, as the inner, slower moving shells successively catch up with the shocked region
and refresh it \citep[][ and references therein]{nakauchi15} . The explosion geometry is is asymmetrical in the former two cases, and 
symmetrical in the case of the refreshed shock. Few radio hypernovae have been detected so far, and again, a radio survey like SKA would be the best 
option to increase the statistics for these events.

Typically,  in SNe, spectral radio luminosity
$L_R \simeq 10^{33}$ to $10^{38}$ erg s$^{-1}$ at 5 GHz, making them fainter by a factor of$ \sim 10^4$ in the radio than in the optical. However, there 
is much diversity in the radio light curves, in the temporal evolution of the radio "spectra", and in the onset of radio emission in different bands, as even 
across the same sub-class, there are both radio-bright and radio-suppressed events. The latter may partly be a selection effect, which could be resolved 
with the high sensitivity of the SKA, using SKA-1~MID bands.

\subsection{Radio supernova searches with SKA}

Radio lightcurves are crucial to understand the history of the progenitor star. 
However, 
currently only $\sim$ 10\% of the discovered
core-collapse SNe show radio emission. 
However, only the targeted searches of some optically bright 
supernovae
have been carried out in radio bands and  systematic searches of radio emission from core-collapse supernovae (CCSNe) are still
lacking.
Optically
dim SNe are missing due to dust obscuration by the host.
Therefore, finding SNe in radio bands are more promising to 
avoid dust obscuration and determine the
true SN rate and thus get a handle on star formation rate. 

With increasing sensitivity of SKA, it will be possible to study more than an order of magnitude SNe in radio bands \citep{perez14}. 
  The
 SKA-1~MID will be able to detect many such dust obscured SNe, 
including those located in the innermost regions of host 
galaxies as radio bands do not suffer from dust extinction. 
The detection rate will provide 
information about the current star formation
 rate as well as the uncertain  initial mass function.
  This will 
enable us to determine the true volumetric 
CCSN rate. SKA-1 is well suited for this.
We would also be able to  
 follow-up bright supernovae for a longer time.  
 Commensal observations along with a weekly cadence is likely to result in the most
complete
 sample of radio supernovae \citep{perez14,wang15}.
With an improved sensitivity level of 1 uJy, one can detect the brightest of radio
SNe, such as the Type IIn SN 1988Z up to redshift of z=1. 


\subsection{Unveiling thermonuclear Supernovae with SKA}

Despite their great cosmological importance, the exact nature of progenitors of SN-Ia remains unknown. While the canonical model is that the SNe itself is caused by thermonuclear explosion of a CO white dwarf (WD), initiated by accretion from a secondary, the nature of the secondary itself is still debatable. In the singly degenerate (SD) progenitor scenario, the H- and/or He-rich secondary could be a non-degenerate star which is either MS, immediately post-MS, or even a late evolved red-giant, while in the double degenerate (DD) scenario it could be another WD in close contact binary, which merges in to the degenerate SNe progenitor, leading to the thermonuclear runaway reaction. Despite the very large number of optical observations of bright, nearby, SNe-Ia, which track the evolution of the SNe from shock outburst to nebular stage, the exact model has not yet been confirmed. Population synthesis studies of SD and DD models \citep{claeys14, rutter09}
suggests that the DD channel is the most probable one, with multiple possible evolutionary sequences, leading to the formation of the close DD system.
In any case, it is to be expected that mass loss would have occurred from the progenitor(s) prior to explosion, via stellar winds, Roche-Lobe overflow (RLOW), or in the common envelope (CE) phase, and a shock-CSM interaction should produce radio emission, with the low frequencies (e.g.) SKA bands reaching peak luminosity at a later time. This implies that the temporal, spectral and flux evolution of the radio emission would serve as a trace of the physical properties of SNe shock and CSM interaction.

  To date, no SNe-Ia has been detected in the radio, though this could also be a selection effect, in that few bright SNe-Ia have been followed up in the radio band, immediately post-shock, and because a less massive and/or dense CSM would mean that the radio flux density would lie below the detection threshold of the current generation of radio telescopes. E.g. for the nearby SN-Ia SN2014J (at 3.5 Mpc), only an upper limits on flux density in the  eMERLIN (37.2 and 40.8 $\mu Jy$ at 1.55 and 6.17 GHz respectively), JVLA (12.0 and 24.0 $\mu Jy$ at 5.5 and 22 GHz respectively) and eEVN (28.5 $\mu Jy$ 1.66 GHz) are known, for epochs spanning up to 35 days post explosion  \citep{perez14} -- this lack of prompt emission appears to be consistent with the DD progenitor model. Thus, for nearby type-Ia SNe ($\simeq 2$ such events are detect each year at $V_{peak}< 13$), even a non-detection of radio emission in the SKA bands would set tight constraints on the mass loss history of the progenitor system, and hence on the theoretical model. It is also to be noted that SNe of all kinds, whose optical emission is shrouded by dust, would nevertheless be visible in the SKA bands, and therefore a survey of transient SKA sources would further constrain supernova rates as a function of various galactic properties.

In addition, Supernova explosion triggers shock wave which 
expels and heats the surrounding CSM and
ISM, so forms supernova remnant (SNR). It is 
expected that more SNRs will be discovered
by the SKA.
This may decrease the great number discrepancy
between the expected and observed.
Several Supernova remnants have been
confirmed to accelerate protons, main component of cosmic rays, to very high
energy by their shocks.
The cosmic ray origin will hopefully be solved by
combining the low frequency (SKA) and very high frequency (Cherenkov
Telescope Array: CTA) bands' observations.

\subsection{Gravitational Lensing of Supernovae}

Time delay between multiple images of a Gravitationally lensed source
  is a powerful diagnostic of the Physics of the source and  Cosmology.
 Time delay in lensing is the only direct observable that gives the physical scale of the system.
Rich Galaxy clusters at intermediate redshift ($\sim$ 0.2 -- 1)
 are powerful lenses. In the central region, multiple images are
 linearly magnified (according to certain characteristic way)
upto a factor of 100, and so, background galaxies at redshift of the order of 1 to 3 are elongated into giant arcs of 10 - 30 arcseconds.
(This is an important channel to possibly detect even galaxies
in forming at redshifts of more than 6)
There are more than 100 highly magnified lensed galaxies recorded
 and the number is increasing with more and more systematic surveys.

If a supernova or GRB occurs in one such galaxy, it will
appear in one of the images first then in the other two images (or three,
if magnification is sufficient for detection)
at time lag of week to year scale. (This was pointed out
by \citet{sw87,nc89}. 
Position of the event  and information
 about which image will have the event seen first can be
unambiguously determined from optical observation of the system
and lens model constructed from it. Multifrequency observations, specially
optical and radio will be valuable in this respect.

Detecting the event in one image (even partially) allows an observer
to plan and monitor the event in other images.
This helps to test many aspects of the event itself with elaborate multifrequency observations.
The subsequent refined models (specifically time delay measurements
are important cosmological probe \citep[specifically Hubble Constant
 and  dark energy, ][]{dev04}.  \citet{metcalf07} pointed out
 that the supernova data favour dark matter made of microscopic particles (such as the lightest supersymmetric partner) over 
 macroscopic compact objects (MCOs).

Since there are more than 100 potential targets,  we expect one year of careful monitor to register one usable event.
SKA-1~MID can organise such a programme and look for these events.

\section{Radio selected transients}

A large fraction of
transients detected in radio for the first time remains undetected in other
bands of electromagnetic spectrum and their nature remains unclassified. 
Early searches for variable and transients in radio band yielded only $\sim
10$ variable sources \citep{gt86,langston90}. 
In the last decade, use of wide frequency imaging  techniques have
resulted in several reported detections. 
These include: (i) detection of a
transient GCRT J1745$-$3009 with flux density $\le 1$ Jy at 325 MHz about
a degree away from the Galactic centre emitting coherent emission \citep{hyman05, roy10},
 (ii) ten transients from multi-epoch (22 years)
observations of a single field from archival VLA data at 4.8 and 8.4 GHz at
$\sim$a few hundred $\mu$Jy or higher level of flux density \citep{bower07}, 
(iii) detection of a single transient at 1.4 GHz with $\sim 1$
Jy flux density using Nasu observatory in Japan  \citep{nii07}, (iv)
detection of a single transient GCRT J1742$-$3001 at 240 MHz with flux density
$\sim 0.1$ Jy near the Galactic centre \citep{hyman09}, (v) detection of
15 transients at 843 MHz from a 22-yr survey with Molonglo observatory
synthesis telescope \citep{bannister11} at $\sim 10$ mJy or higher, (vi)
detection of a single transient from about 12 hours of observation at 325 MHz
at a few mJy level \citep{jaeger12}, (vii) detection of a single
transient from 4 months of observations at 60 MHz with LOFAR at $\sim 10$ Jy
level \citep{stewart}.

Out of the 30
transients detected, possible progenitor population of only about 9 have been
identified by the respective authors. The origin of the rest of
the transients are not known. Clearly, this demonstrates that for most of the
transients detected first in radio band, the parent population is unknown.
If we consider extragalactic origin for the transients of unknown origin, we
note that flux densities of the detected ones except in \citet{bower07}
are significantly higher than expected from orphan gamma-ray burst afterglows
(mJy or lower). Given their flux densities, most of these sources could not be
GRB counterparts in radio. Considering optical images of their fields and flux
densities, most of these sources could not also be RSNes. Lack of optical
emission in continuum images also argues against any scintillating AGN
scenario. Therefore, most of the above transients do not appear to have known
type of extragalactic progenitor source.

Transient emission could occur from sources within our Galaxy. However, the
flux densities of the sources detected in radio either lie above or near the
upper limit of known stellar luminosity distribution in radio \citep{gudel02,bower07}. 
Therefore, typical stellar radio emission cannot account
for these emissions. However, coherent emission from stellar surface could
occur from energetic processes (e.g., re-connection of magnetic flux tubes)
like the electron cyclotron maser at the electron gyro-frequency and its
harmonics, or the plasma emission at plasma frequency or its lower order
harmonics. Though the bandwidth of such emission is quite small, but there can
be multiple spots of emission with slightly different centre frequency \citep{roy10}. 
These emission could have high brightness temperature reaching
$\sim 10^{16} - 10^{20}$ \citep{ergun00,stepanov01}. These
types of flare emission from late type stars (e.g., M-type Dwarfs) is known
and detected in radio. Their flux densities recorded at cm wavelengths are
typically in millijanskys. However, at metre wavelengths or larger, some of
the flares recorded are $\sim$100 Jy \citep{aa95}. Therefore,
occasional radio emission from magnetically dominated flare stars which are
quite common in the Solar neighbourhood could explain some of these transient
emission \citep{bower07,roy10,jaeger12,stewart}
 and could dominate the transient emissions at metre or decameter
wavelengths.


Lack of identification of origin of a major fraction of radio detected
transients at the present era makes it a challenging problem. SKA-1 will
have a collecting area a few times larger than the GMRT, which will allow to
probe for day time scale transients down to $\sim 10 \mu$Jy within a bandwidth of a few tens of MHz at metre
wavelengths. More importantly, at that wavelength range, the SKA~LOW will have
a large field of view, which will allow to observe a significant fraction of
the sky at any time. Detecting long time scale weak transients will not be
possible in real time. However, if near real time images of the sky are made
with automated pipelines on daily basis, then a comparison of the same portion
of the sky observed earlier will allow to detect these transients. This could
trigger quick followup observations at other wavebands, which will help to
identify their progenitors. Also, a better sensitivity would ensure detection
of the brighter ones while their flux density rises/decays down. This will
provide their rise/decay light curve crucial to compare with known type of
transients. From the earlier detection of transients by \citet{bower07} at
5 GHz and \citet{bannister11} at 843 MHz (assuming a steep spectral index
of 0.7 for the transient flux densities, and no. density of transients to fall
down as a power of flux density with an index of $- 1.5$) , and a cutoff flux
density of 0.1 mJy at 330 MHz (10$\sigma$) for the SKA-1, we estimate
the rate of detection of long timescale transients to be $\sim 1 - 2$ per
sq-deg per day. This detection rate is consistent with \citet{jaeger12}, but almost an
order of magnitude lower than what is expected from \citet{stewart}
result at 60 MHz scaled to the above flux density limit at 330 MHz. However,
in the last two works, the authors discovered only one transient in each case,
and only upper limit can be derived from their results.

It should be noted that the above estimation of the rate of transients
assumes observations far away from the Galactic plane. Results from
observations near the central region of the Galaxy have resulted in detection
of 4 transients with GMRT and VLA \citep{hyman02,hyman06,hyman07,hyman09}.
 These
observations were done over $\sim$30 epochs in the last one decade and
detection threshold was above a flux density limit of 0.1 Jy. This detection
rate (0.005 per sq-deg per day) is an order of magnitude higher than is
estimated from above. This does show a Galactic contribution of transients in
radio, the progenitor population of which needs to be identified. This will
also help in separating the two types of transients based on their location in
our Galaxy or external ones.

\section{Novae Outbursts}

Novae form an important sub-class of transients in our Galaxy as well as in
other galaxies. A nova outburst is triggered by runaway thermonuclear burning 
on the surface of an accreting white dwarf in an interacting binary system. 
With outburst energy in the range of $10^{38}$ to $10^{43}$ ergs, these events
are amongst the energetic explosive ones. At an estimated rate of 30 novae per
year (2002), these events are fairly frequent. Although the current observed 
rate is much lower than estimated, future deep surveys by facilities such as
the LSST are expected to increase the number of observed events. Nova systems 
serve as valuable astrophysical laboratories in the studies of physics of 
accretion onto compact, evolved objects, and thermonuclear runaways on 
semi-degenerate surface which give insight into nuclear reaction networks. They
also contribute to enriching the ISM with heavy isotopes of $^{13}$C, $^{15}$N,
$^{22}$Na and $^{26}$Al. In addition, if the mass ejected during a nova outburst
is less than that accreted since the last outburst, there is the possibility of
the WD growing in mass, making (at least some) of these systems interesting
candidates for SNe Ia progenitors. However, despite their astrophysical 
significance as nearby laboratories, many aspects of these relatively common 
stellar explosions remain poorly understood. Although much of our current 
understanding of these systems has come from the optical observations, 
multi-waveband observations have augmented and enhanced our understanding. 
Radio observations play a key role in addressing some of the puzzling aspects 
of accretion, outburst and interaction with the environment. 

Radio emission from novae typically lasts longer than the optical emission, on 
timescales of years, rather than months. Observations at different epochs yield
information on different aspects of the nova outburst. While the very early 
phase observations provide information regarding the distance to the nova, as
the nova evolves, the observations provide clues to the mass of the ejecta. 
The mass of ejecta is a fundamental prediction of nova models, and thereby 
provides a direct test of nova theory. The primary mechanism of radio emission
is thermal bremsstrahlung from the warm ejecta. In addition, in the case of
novae in dense environment, the interaction of the nova ejecta with the
environment gives rise to a shock, which in turn may give rise to a non-thermal,
synchrotron emission component, like in the case of GK Per \citep{sequist89, ak05}, 
RS Oph \citep{hjel86, taylor89, obrien06, kantharia07, rupen08, soko08, eyres09}
and V745 Sco \citep{kantharia16}. The {\it Fermi} 
detection of recent novae such as V407 Cyg 2010, V959 Mon 2012, V1324 Sco 2012, 
V339 Del 2013 and V745 Sco has revealed that shock interaction with the dense 
environment can also lead to GeV gamma ray emission. 

Previous radio observations of nova outbursts have illustrated the unique 
insights into nova explosions these observations bring. Since radio emission 
traces the thermal free-free emission, by extension it traces the bulk of the 
ejected mass. In addition, radio observations are not subject to the many 
complex opacity and line effects that optical observations both benefit and 
suffer from. Thus, in addition to being relatively easier to interpret, radio 
observations can also probe how the ejecta profile and dynamic mass loss evolve 
with time. Radio observations of novae will thus illuminate the many 
multi-wavelength complexities observed in novae and test models of nova 
explosions.

SKA and its precursors will be very well placed to  study novae at various epochs of the outburst. While the thermal
emission from most novae have been well observed in the higher ($>1$ GHz) 
frequencies, the sensitivities of the existing facilities are not well suited to
detect thermal emission at $<1$ GHz. The improved sensitivity of SKA at the
lower frequencies will enable detection of the thermal emission at these
frequencies, providing a better understanding of the evolution of the 
physical conditions in the nova ejecta. Also, the $<1$ GHz frequencies are ideal
to observe the non-thermal emission from novae, especially from the recurrent
nova systems \citep[see ][]{kantharia16}. A sensitivity limit of 1 mJy can 
detect radio emission at $<1$ GHz upto a distance of 10 kpc, if the non-thermal 
luminosity of the nova system is $10^{13}$ W~Hz$^{-1}$ \citep{kantharia12}. An 
important motivation for studying the non-thermal radio emission from recurrent 
novae is to interrogate possible evolutionary connection to the lack of 
detectable radio emission from type Ia supernova systems. As has been shown by
\citet{kantharia16} for the recurrent novae RS Oph and V745 Sco, 
changes are observed in circumstellar material with subsequent outbursts. Observations of
future outbursts will establish if these changes are evolutionary, and also 
their impact on the evolution of the binary system itself.

\section{Multi-messenger Astronomy with SKA: Gravitational Wave perspective}

Electromagnetic counterparts of gravitational waves (GWs) are another important class of transients relevant to SKA science.
Laser Interferometric Gravitational wave observatory (LIGO) recently reported the first detection of a binary black hole merger \citep{abott16a},
 opening a new observational window to the universe through GWs.

This binary  black hole event (named as GW150914) was followed up electromagnetically \citep{abott16} and in neutrinos \citep{atri16} using various observational facilities in the world. While it has been argued that BH-BH mergers will not produce 
electromagnetic counterparts \cite{gw16}, and indeed no electromagnetic counterparts associated with GW150914 were reported, this sets milestone in the field of multi-messenger astronomy where a real GW signal was followed up various electromagnetic (EM) bands. It is worth noting that the above mentioned follow ups included radio follow ups by LOFAR, MWA and ASKAP which demonstrates the capabilities of radio telescopes to carry out searches for EM counterparts of GW events. The LIGO configuration which lead to the detection of GW150914 is going to evolve to much better sensitivities which will bring the detection of NS-NS and NS-BH binaries within its reach and more GW detectors are going to come online in the next few years which will considerably improve the source localization.  

Phenomenon such  as short GRBs and (a sub-population), FRBs have been proposed to be associated with NS-NS or NS-BH mergers. The science potential ranges from detecting radio afterglows associated with short GRBs coincident with GW observations of NS-NS or NS-BH binaries (hence directly confirming the compact binary progenitor of short GRBs), detecting orphan afterglows associated with some of the NS-NS or NS-BH mergers which might have produced a GRB which was pointed away from us and hence missing it, to somewhat speculative scenario of FRBs being detected in coincidence with GW observations (if a sub-population of the FRBs is produced due to mergers of NSs). Such joint observations can shed light on various aspects of the phenomenon including possible constraints on the models of GRB jets \citep{arun14}.

One of the interesting prospects of SKA for multi-messenger astronomy lies in the radio follow up of GW triggers using the second or third generation gravitational detectors detectors that may be operational at the time of SKA.  While it has been argued that BH-BH mergers will not produce 
electromagnetic counterparts \cite{gw16}, there are enough uncertainties in the electromagnetic science of GW waves. In addition, the network of GW detectors by the time SKA comes online, will have much higher sensitivity to detect NH-NH, NH-BH mergers, and may have the capability to localize the source to within few square degree which makes efficient follow up of the source much simpler. The FOVs of ASKAP (30 sq degree) and SKA~LOW (27 sq degree)  \citep[see Table 1 of ][]{ska} suits requirements for the EM follow up of GW triggers using a network of interferometers which will localize the sources within a few sq degrees to a few tens of sq degrees (see Table 3 of \citet{fair11} for instance), depending on the nature of the compact binary. The very fact that ASKAP took part in the radio follow up search of GW150914, shows the preparedness of SKA-like detectors to take part in such follow up efforts. The science returns from such joint radio-GW observations are enormous. 
It is clear that ASKAP will continue to play a crucial role in the follow up of GW triggers in the upcoming  science runs of LIGO.
It is interesting to note that SKA will be one of the unique instruments which by itself will be searching for Gravitational Waves (from supermassive BH binaries)  in the pulsar timing array mode and also looking for electromagnetic counterparts to ground-based GW detector triggers. Hence multi-messenger astronomy associated with these systems is going to be very exciting.

\begin{figure}
\centering
\includegraphics[width=0.45\textwidth,natwidth=610,natheight=642]{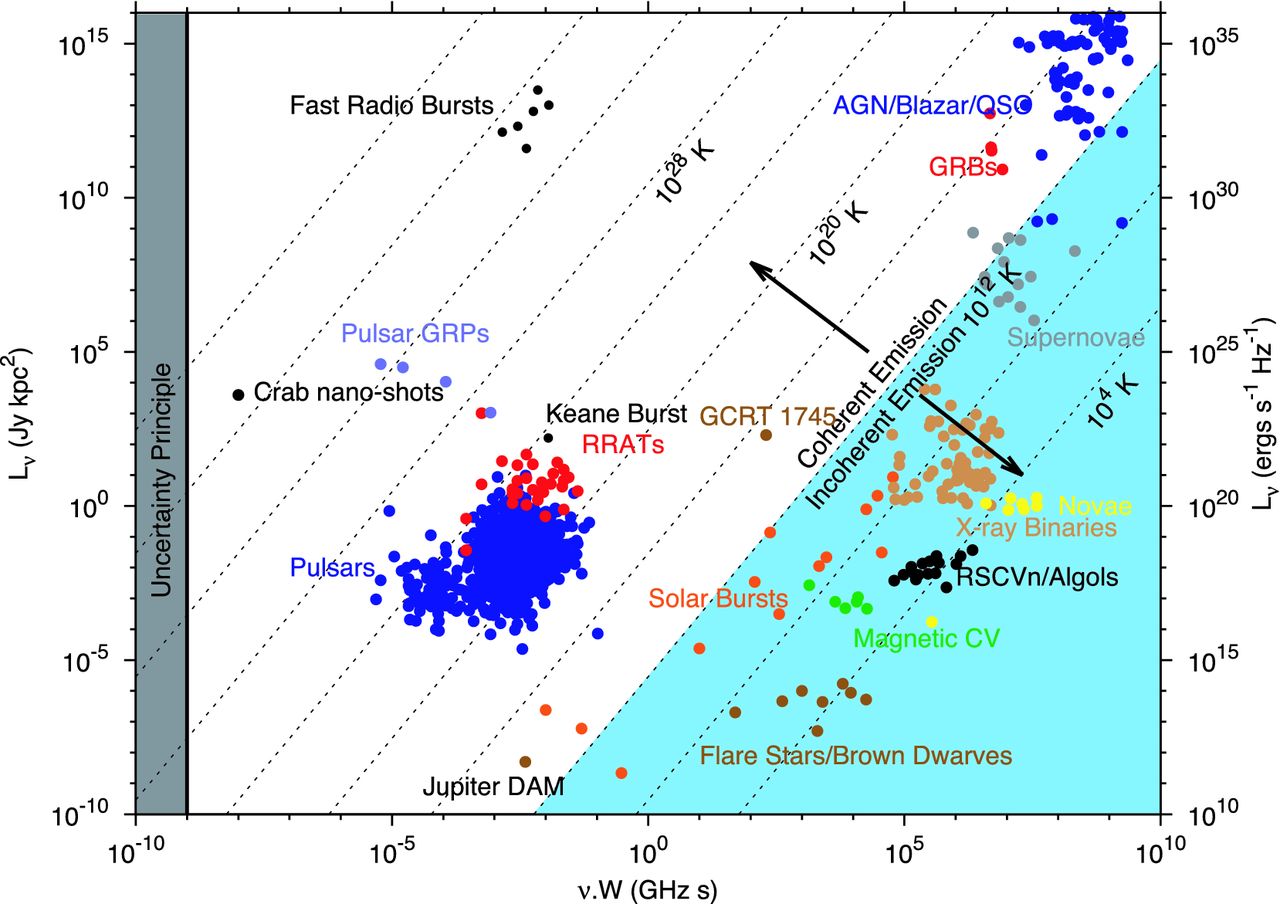}
\caption{Parameter space for transient sources. This allows one to identify the sources of coherent radio emission (pulsars, fast radio bursts, certain emission from Jupiter and the Sun, etc.) for comparison with the more slowly varying synchrotron transients \citep[reproduced from ][]{fpk15}.}
\label{fig4}
\end{figure}

\section{Discussion and Conclusions}

With the high sensitivity and wide-field coverage of the SKA, very large
samples of explosive transients are expected to be discovered. A
major fraction of these will be tidal disruption events (TDE)  followed by type II SNe and
orphan afterglows of GRBs and relativistic supernovae. Even at
the ASKAP (precursor) stage, VAST-Wide surveys conducted at 1.2 GHz band
for about 10,000 square degrees every fortnight will typically be
expected to yield some 32 type II SNe, 82 {\it Swift} 1644+57 TDE events, 8 orphan
afterglows \citep{frail2012}. The large number of such newly discovered
sources will be a rich harvest, among which there will be quite a few that
will have characteristics (such as radio brightness, counterparts at other
wavebands or other physical characteristics like outflow properties) that
will make it possible to follow them up intensively with high cadence. Not
only will these new discoveries
trace out well-populated areas of phase space of explosions see Fig. \ref{fig4},
but quite likely, they will provide hitherto unknown linkages that will clarify
or strengthen suspected unity among diversity \citep{fpk15}. 

Radio wavelengths are 
particularly well suited for uncovering such phenomena in synoptic surveys, 
since observations
at radio wavelengths may suffer less obscuration than in other bands (e.g.
optical/IR or X-rays) due to dust and other absorption. 
At the same time  a multiwaveband
approach is a "force-multiplier", since source identification becomes
more secure and multiwaveband information often provides critical
source classification rapidly than possible with only radio band data.
Therefore,
multiwaveband observational efforts with wide fields of view will be the
key to progress of transients astronomy from the middle 2020s offering
unprecedented deep images and high spatial and spectral resolutions.
Indian astronomers with wide-ranging experience of low frequency radio
astronomy in a variety of astronomical phenomena and targets would be 
particularly well-placed to pursue time critical transient objects 
with SKA and observatories at other bands.

Our strategy will also include  to develop tools with machine learning
capabilities to automatically and rapidly classify the
transient events expected in the SKA era and follow up a few of these
intensively and with multiwaveband coverage through other large facilities
that India will likely have access to. These exceptional few objects
have the potentials for discovery of major underlying physical processes
and trends. For example the abundance of relativistic SNe without 
GRB counterparts, and luminous TDE events will have implications among
other things on the
sources of Ultra High Energy Cosmic Rays within the
Greisen-Zatsepin-Kuzmin (GZK) cutoff distance from the Earth \citep{chakra11}.

The SKA will be a premier instrument for transient science. This strength of the
science case will  continue to increase as more and more class of transients are
discovered with current surveys, including uGMRT, ASKAP, MWA etc. 
There is much development going on in hardware, software, simulation and data analysis techniques, 
all to improve the chances of detecting transients. All of the next generation telescopes 
telescopes have included the transient science as one of the core science goals, and this is also being reflected in developments
 at nearly all other wavelengths. 

\section{Acknowledgements}

P. C. acknowledges support from the Department of Science and Technology via SwaranaJayanti Fellowship award (file no.DST/SJF/PSA-01/2014-15).

\end{document}